%% file: main.tex
\documentclass[a4paper]{article}

\usepackage{INTERSPEECH2019}
\usepackage{enumitem}

\title{Coarse-to-fine Optimization for Speech Enhancement}
\name{Jian Yao and Ahmad Al-Dahle}
\address{
  Apple Inc., United States of America}
\email{\{jian\_yao, ahmad\}apple.com}

\begin{document}

\maketitle

\input{abstract}

\input{intro}

\input{approach} 

\input{exp}
\input{conclusion}

\clearpage
\bibliographystyle{IEEEtran}
\bibliography{mybib}


\end{document}

%% file: abstract.tex
\begin{abstract}
In this paper, we propose the coarse-to-fine optimization for the task of speech enhancement. Cosine similarity loss \cite{cosineSim} has proven to be an effective metric to measure similarity of speech signals. However, due to the large variance of the enhanced speech with even the same cosine similarity loss in high dimensional space, a deep neural network learnt with this loss might not be able to predict enhanced speech with good quality. Our coarse-to-fine strategy optimizes the cosine similarity loss for different granularities so that more constraints are added to the prediction from high dimension to relatively low dimension. In this way, the enhanced speech will better resemble the clean speech. Experimental results show the effectiveness of our proposed coarse-to-fine optimization in both discriminative models and generative models. Moreover, we apply the coarse-to-fine strategy to the adversarial loss in generative adversarial network (GAN) and propose dynamic perceptual loss, which dynamically computes the adversarial loss from coarse resolution to fine resolution. Dynamic perceptual loss further improves the accuracy and achieves state-of-the-art results compared with other generative models.
\end{abstract}

%% file: intro.tex
\section{Introduction}

Speech enhancement aims at improving the quality of the speech contaminated by the additive noise. It has quite a few applications including noise cancelling, audio editing, preprocessing for speech recognition, just to name a few. Denote the noisy speech as $y(t)$, we have
\begin{equation}
y(t) = x(t) + n(t)
\end{equation}
where $x(t)$ and $n(t)$ are respectively the clean speech and the noise, with $t$ being the time index.
Speech enhancement tries to recover the clean speech $x$\footnote{For simplicity, we omit the time index for the audio signal.} from the noisy speech $y$. Traditionally, Spectral Subtraction\cite{specsub} and Wiener Filtering\cite{wiener} are two popular speech enhancement algorithms. Spectral Subtraction approach needs to estimate the noise spectrum and subtracts it from the noisy speech. However the characteristics of the noise is not trivial to approximate, and the noise spectrum might not be separable from the clean speech in frequency domain. Wiener Filtering instead tries to recover the clean speech by estimating the ratio between the spectrum of the clean speech and the noisy speech. The inaccurate approximation of the ratio limits the wide application of Weiner Filtering in practice.
Recently, deep learning based algorithms have shown promising results in speech enhancement\cite{segan, deepfeat, mgan, rnnse, phase, waveform_utter}. These approaches can be further categorized into generative models\cite{segan, mgan, add1, add2} and discriminative models\cite{reg, multinoise, lstmEN, autoencoder}. In discriminative model, the deep neural network (DNN) takes the noisy speech as the input and tries to predict the clean speech, i.e. DNN is directly modeling the conditional distribution $p(x|y)$. 
While in generative model, conditional GAN \cite{gan, cgan} is the prevalent method for speech enhancement, which models the distribution of $p(x, z|y)$, with the additional variable $z$ being latent variables. 
There are two key components in GAN, which are respectively called the generator $G$ and the discriminator $D$. 
During the training, the parameters of the generator are tuned so that its prediction can fool the discriminator. In the meanwhile, the discriminator also evolves to be more capable of differentiating between the synthetic data from the generator and the real data. The training process can be written as follows:
\begin{align}
\min_D V(D) = & -\mathbb{E}_{x,y \sim p_{data}(x, y)}[logD(x, y)] \label{dloss}\\
              & + \mathbb{E}_{z \sim p_z(z), y \sim p_{data}(y)}[log(D(G(z, y), y))] \nonumber \\
\min_G V(G) = & -\mathbb{E}_{z \sim p_z(z), y \sim p_{data}(y)}[logD(G(z, y), y)] \nonumber\\
              & + \mathbb{E}_{x, y\sim p_{data}(x,y)}[\mathcal{T}(G(z, y), x)] \label{gloss}
\end{align}
where $\mathcal{T}$, usually called regularization function, is a traditional loss function such as $L1/L2$ loss or cosine similarity loss.
Both \cite{mgan} and \cite{segan} have shown that the GAN approach works well only if the traditional loss term $\mathcal{T}$ is added in Eq.~\ref{gloss}. This observation has also been confirmed in image synthesis\cite{pixel2pixel} using conditional GAN, where the training example is a pair of images instead of audios. Note that $\mathcal{T}$ can be by itself used as the loss function in discriminative models. The training of a discriminative model can be written as follows:
\begin{equation}
\min_w \mathbb{E}_{x, y\sim p_{data}(x,y)}[\mathcal{T}(f_w(y), x)]
\label{disc_eqn}
\end{equation} 
where $f_w$ is the function parametrized by $w$ that maps noisy speech $y$ to the enhanced speech $\hat{x}$. 
Comparing Eq.~\ref{disc_eqn} with Eq.~\ref{gloss}, there is an additional term in Eq.~\ref{gloss} which is the adversarial loss. In this regard, learning the generator in GAN can be seen as the training of a discriminative model with dynamic loss as the parameter of the adversarial loss keeps changing during training. Here the effect of $z$ is neglected, as the one to one mapping from $y$ to $x$ can still be learnt without $z$ in GAN, producing deterministic outputs. The stochasticity of the prediction from the generator is still an area of research, particularly in the one-to-one mapping problem\cite{pixel2pixel, z1, z2}. 

In most of previous work, the loss function $\mathcal{T}$ is either computed by aggregating the $L1$/$L2$ distance\cite{segan, cnnse, rnnse} for each component of the audio or computed by evaluating the entire predicted audio as a whole in high dimensions such as cosine distance loss\cite{cosineSim, phase}.   
However, as far as we are aware, there is no work in speech enhancement that optimizes the loss from coarse to fine in training. Coarse-to-fine strategy first optimizes the loss on the high dimensional audio sequences and then gradually reduces the granularity of the evaluated sequence to compute the loss. It not only takes advantage of fast convergence from coarse granularity but also reduces saturation and fine tunes the details as the granularity gets finer\cite{c2f1, c2f2}. Thus the contributions of this paper are listed as follows:
\begin{itemize}[leftmargin=0.4cm]
\item We propose a general coarse-to-fine optimization for speech enhancement which can be applied to both generative and discriminative models. 
\item We extend the idea of coarse-to-fine to the adversarial loss in the training of the generator of GAN and propose dynamic perceptual loss.
\item Experimental results show that the coarse-to-fine optimization outperforms a single granularity in quantitative metrics. Meanwhile, the proposed coarse-to-fine optimization and dynamic perceptual loss achieve the new state-of-the-art for both discriminative models and generative models.
\end{itemize}

%% file: approach.tex
\section{Basics}
\subsection{Cosine Similarity Loss}
Cosine similarity loss\cite{cosineSim} is widely used to measure the similarity between two vectors, which is defined as:
\begin{equation}
\mathcal{T}(\hat{x}, x) = -\frac{\hat{x}^T x}{||\hat{x}||_2 ||x||_2}
\label{eqncos}
\end{equation}
It is chosen in our algorithm because 1) it shows better accuracy than $L1/L2$ loss even with a single granularity. 2) it can be evaluated in different granularities, as opposed to $L1/L2$ loss which is essentially computed by aggregating the $L1/L2$ loss in every single dimension of $x$ (the finest granularity).
As shown in Eq.~\ref{eqncos}, cosine similarity loss function actually computes the cosine value of the angle between two vectors. As the dimensionality of the vector increases, with the same cosine similarity loss between the prediction $\hat{x}$ and the ground truth $x$, the number of feasible solutions of $\hat{x}$ also increases which adds uncertainties to the prediction. Therefore, if we optimize Eq.~\ref{eqncos} in different granularities from high dimension (coarse) to low dimension (fine), the resulting prediction will be more constrained and better resemble the true audio sequence.
\subsection{Time Domain vs Frequency Domain}
Either the waveform or the spectrum of the audio can be the input and the output of the neural network. They are in essence the same as the Fourier Transform and Inverse of Fourier Transform can be represented by convolutional layers with fixed weights. However, in practice we observed that training directly from raw waveform takes longer than training from spectrum to converge to a reasonably good result. We also tried to encode the Fourier Transform as the convolutional layer in the DNN and initialize the corresponding layer with Fourier Transform coefficients but allow those coefficients to be further fine tuned with input and output being both raw waveform. We found the resulting accuracy is not better than the case where we just use fixed Fourier Transform coefficients. Furthermore, the training is observed more stable in frequency domain than that in time domain. Occasionally, the training in time domain could end with unacceptably poor results. 
Note that even the spectrum is used as the input/output, the loss can still be computed in time domain and back-propagation can be used to train the network because of the fact that all operations in inverse of Fourier Transform are differentiable.
Thus in this paper, the DNN which maps the noisy speech $y$ to the enhanced speech $\hat{x}$, is learnt in frequency domain. More precisely, we first apply Short Time Fourier Transform (STFT)\cite{stft} to the noisy speech and obtain its spectrum. Then instead of directly predicting the ground truth spectrum from the input spectrum, the network will predict a complex-valued mask of ratio constrained by a $tanh$ function\cite{cmask}. The spectrum of the predicted audio $\hat{x}$ can be simply derived by multiplying the input spectrum with the mask. By using the complex-valued mask as the output of the DNN, not only the magnitude but also the phase information is restored from the noisy input audio. In inference, the spectrum will be converted to the waveform via inverse STFT. In training, the cosine similarity loss will be computed on the waveform, following the inverse STFT. In all our experiments, the time window of STFT is 1024 with stride 256. The number of bins in frequencies is effectively 513 as the Fourier Transformation is applied on real numbers. 

\begin{figure}[t]
  \centering
  \includegraphics[width=0.6\linewidth]{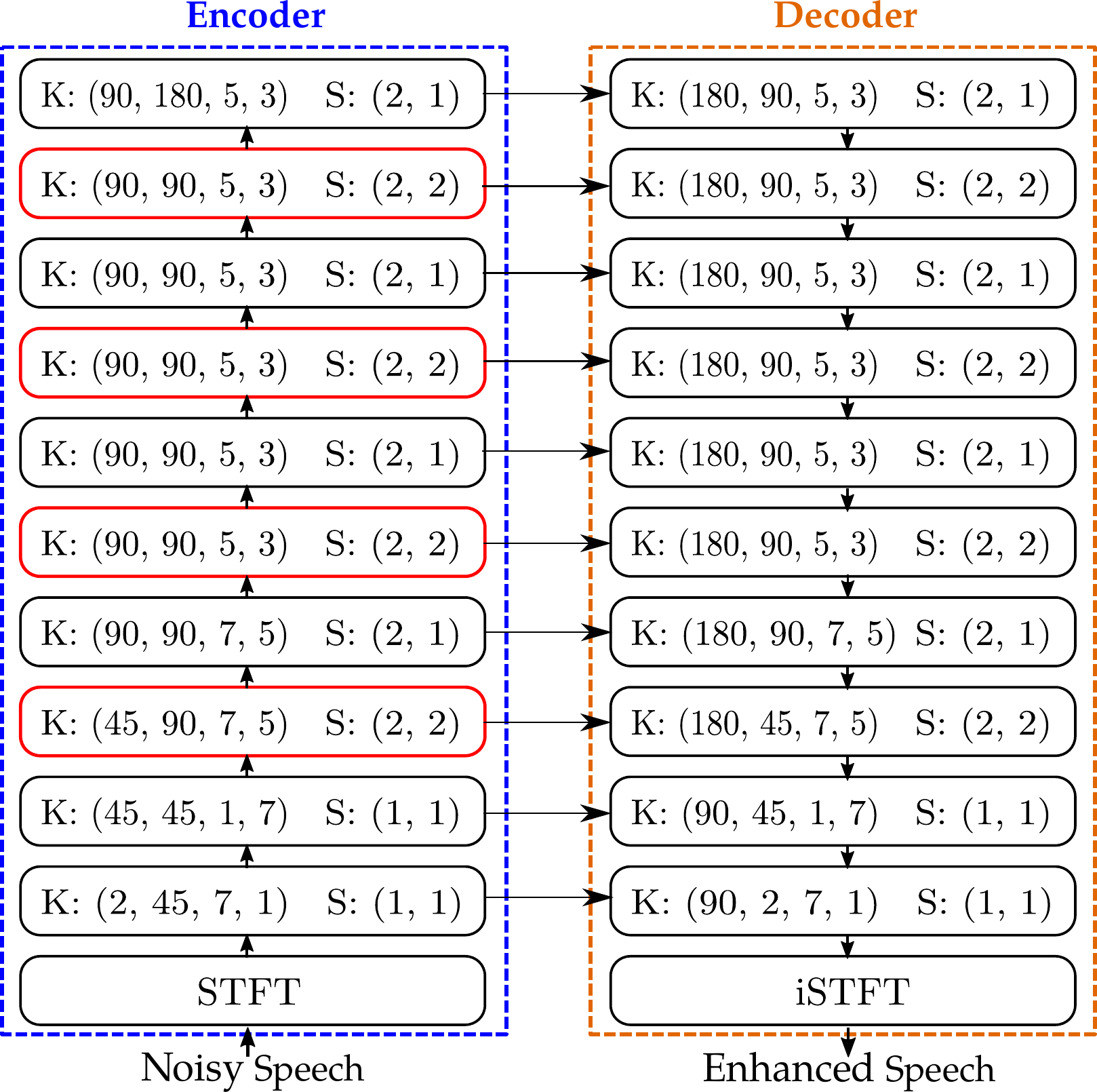}
  \vspace{-1mm}
  \caption{The network architecture of our discriminative model and the generator of our generative model. Each block consists of convolution/deconvolution\cite{deconv}, batch norm\cite{bn} and leaky RELU\cite{lrelu}. The kernel shape is denoted as $K$: (in\_channel, out\_channel, k\_height, k\_width) and the stride is denoted as $S$: (s\_height, s\_width). The red blocks are layers 9, 7, 5, 3 from top to bottom, where dynamic perceptual loss is computed. \vspace{-5mm}}
  \label{fig:network_arch_gnet}
\end{figure}
\subsection{Network Architecture}
\begin{figure}[t]
  \centering
  \includegraphics[width=0.6\linewidth]{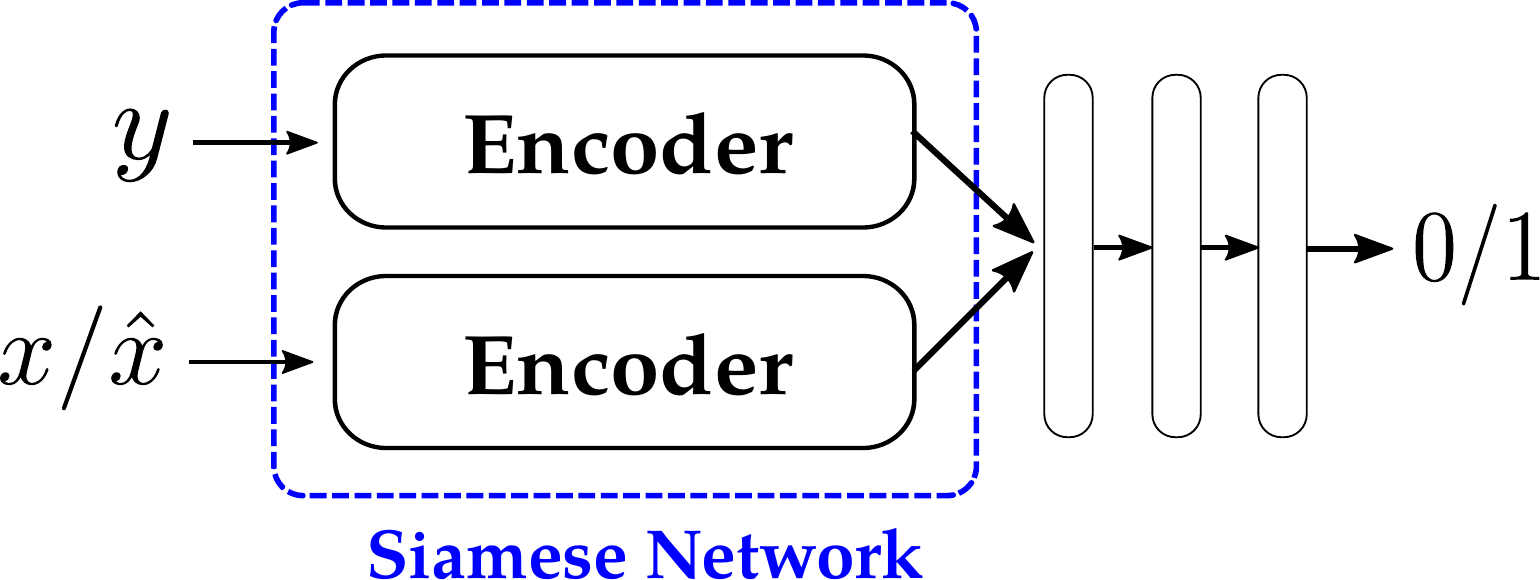}
  \vspace{-1mm}
  \caption{The network of the discriminator of our GAN model. $y$: noisy speech. $x/\hat{x}$: clean speech$/$enhanced speech.\vspace{-3mm}}
  \label{fig:network_arch_dnet}
\end{figure}

Encoder-Decoder style network is popularly used in both generative model\cite{segan} and discriminative model\cite{cnnse, phase}. The network architecture in our approach also follows the encoder-decoder style, enabling our proposed approach to be directly compared with other methods. For the discriminative model, we use the same network architecture as cRM$\mathbb{R}$n with 20 layers in \cite{phase}, a state-of-the-art discriminative approach, to produce the enhanced speech. Fig.~\ref{fig:network_arch_gnet} visualizes the structure of this network. For our generative model (conditional GAN), the generator will be the same as the network in our discriminative model while the discriminator takes the encoder part of the generator as the backbone, followed by 2 additional convolution layers and one fully connected layer.
Since the discriminator in conditional GAN would take both the data to be classified and the conditional data as input, the backbone of the discriminator will act as a siamese network\cite{deepface} and the additional convolutional layers and fully connected layers in our discriminator will further combine them and classify whether the input data is fake or real. With the same notation as in Fig.~\ref{fig:network_arch_gnet}, the kernel shapes of additional convolution layers are respectively $(360, 512, 3, 5)$ and $(512, 512, 1, 1)$ without batch norm and the stride is $1$. The output of fully connected layer is a one-dimensional scalar. One of the benefits from siamese network is to efficiently compute so-called {\bf dynamic perceptual loss}, which will be discussed in the next section. Fig.~\ref{fig:network_arch_dnet} shows the architecture of the discriminator network in our generative model.

\section{Coarse to Fine Optimization}
The speech signal $x$ can be seen as either a single high dimensional vector or the concatenation of multiple low dimensional vectors. The size of the vector decides the granularity how we divide the signal $x$. As mentioned earlier, computing $L1/L2$ loss in different granularities makes no difference as $L1/L2$ loss is evaluated on every component (finest granularity) of $x$ and then averaged. However the granularity matters when cosine similarity loss is computed!
Denote the dimension of vectors of different granularities as $\{g_i = \frac{dim(x)}{K_i} | i=1, \cdots, n \}$ and $K_a < K_b$ if $a < b$, where $K_i$ is the number of vectors of $i$\textsuperscript{th} granularity and $g_i$ is the dimension of the vector of the corresponding granularity. When $i=1$, the granularity is the coarsest and the dimension of that granularity is just the dimension of the original signal $x$ if we set $K_1 = 1$. While $i = n$, the original vector is divided into low dimensional vectors of the finest granularity.

In the training, we optimize the loss function from coarsest granularity to finest granularity. In each granularity, the loss function can be written as
\begin{equation}
\frac{1}{K_i}\sum_{m=1}^{K_i}\mathcal{T}(\hat{x}[(m-1)g_i : m g_i], x[(m-1)g_i : m g_i])
\end{equation}
\vspace{-0.5mm}where $\mathcal{T}$ is defined in Eq.~\ref{eqncos} and $x[s:r] = \{x[s], x[s+1], \cdots, x[r-1]\}$ is the slicing operation. The optimization for a particular granularity completes when the change of the loss is small or the number of iterations reaches a maximum. In this way, we can reduce the variance when the loss is computed only in high dimension and better resemble our predicted audio to the ground truth. In discriminative models, the coarse-to-fine optimization can be directly applied in the training. While in generative models, this strategy can also be deployed if we use cosine similarity loss as the regularization term in Eq.~\ref{gloss}. 

The coarse-to-fine strategy is not limited to optimize the cosine similarity loss. It can be extended to the problems where the objective to be optimized is not in fine granularity, but the prediction requires finer granularity. Following this philosophy, we propose dynamic perceptual loss. In GAN, the discriminator only gives a confidence score indicating whether the input audio is fake or real. In practice, when we optimize the generator by minimizing Eq.~\ref{gloss}, the first term enforce the prediction from the generator to be like real, i.e. the confidence score from the discriminator to be 1. However, a single score might not be enough to supervise the generator to generate a 'real' audio as the spatial information is lost. Instead, the dynamic perceptual loss will not only enforces the confidence score in the discriminator, but also the deep features with different resolutions in the intermediate layers to be similar to the real audio. Denote the deep feature map in layer $l$ as $D_l(\cdot)$ and the loss as $\mathcal{L}$. The dynamic perceptual loss ($DPL$) in layer $l$ can be written as \vspace{-0.5mm}
\begin{equation}
DPL_l = \mathcal{L}(D_l(G(z, y)), D_l(x))
\end{equation}
\vspace{-0.5mm}Since the siamese network is used in the discriminator of our generative model, it is straightforward to use it for computing both the deep features as well as the confidence score within the same network. Note that if we use other networks instead of siamese network like the one used in SEGAN\cite{segan}, the conditional data and the data to be classified are concatenated at the beginning of the network, which prevents us from computing the deep feature either for real audio or fake audio without concatenating the conditional data.

%% file: exp.tex
\begin{table}[t]
  \caption{Comparison with Other Methods: For Deep Loss, the numbers in parentheses are what we got from their open source code. \textbf{D}: discriminative model which is our own implementation of cRM$\mathbb{R}$n\cite{phase} for fair comparison. \textbf{D+M}: Coarse-to-fine Optimization of \textbf{D}. \textbf{G}: GAN with single granularity optimization. \textbf{G+M}: GAN with coarse-to-fine optimization of \textbf{G}. \textbf{G+M+P}: Dynamic perceptual loss on top of \textbf{G+M}.\vspace{-1.5mm}}
  \label{tab:comp}
  \centering
  \resizebox{0.45\textwidth}{!}{
  \begin{tabular}{ l c c c c c }
    \toprule
              & \textbf{CSIG} & \textbf{CBAK} & \textbf{COVL} & \textbf{PESQ} & \textbf{SSNR} \\
    \hline
    Wiener\cite{wiener}       & 3.23          & 2.68          & 2.67          & 2.22          & 5.07          \\
    SEGAN\cite{segan}         & 3.48          & 2.94          & 2.80          & 2.16          & 7.73          \\
    WaveNet\cite{wavenetSE}   & 3.62          & 3.23          & 2.98          & N/A           & N/A           \\
    MMSE-GAN\cite{cmask}      & 3.80          & 3.12          & 3.14          & 2.53          & N/A           \\
    Deep Loss\cite{deepfeat}  & 3.86(3.79)    & 3.33(3.27)    & 3.22(3.14)    & (2.51)        & (9.86)        \\ 
    \hline        
    D\cite{phase} & 3.79       & 3.32          & 3.20          & 2.62          & \textbf{9.90}          \\
    D+M          & \textbf{3.94}       & \textbf{3.35}          & \textbf{3.33}          & \textbf{2.73}          & 9.40          \\ 
    \hline
    G            & 3.83       & 3.27          & 3.20          & 2.57          & 9.36          \\
    G+M          & 3.94       & 3.33          & 3.31          & 2.67          & \textbf{9.50}          \\
    G+M+P        & \textbf{4.00}       & \textbf{3.34}          & \textbf{3.34}          & \textbf{2.69}          & 9.40          \\
    \bottomrule
  \end{tabular}}
  \vspace{-6mm}
\end{table}
\section{Experiments}
In this section, the experimental results show that in either discriminative models or generative models, the coarse-to-fine optimization will improve the current state-of-the-art algorithms. Particularly in generative models, the proposed dynamic perceptual loss could further improve the accuracy obtained from optimizing the cosine similarity loss from coarse to fine.
\subsection{Dataset and Metrics}
\subsubsection{Dataset}
\vspace{-1mm}
We evaluate our proposed algorithm on speech enhancement dataset by Valentini et al.\cite{dataset}, which is widely used in other popular speech enhancement methods\cite{segan, mgan, deepfeat, phase}. This dataset consists of 11572 mono audio samples for training and 824 mono audio samples for testing. The duration of the audio ranges from 1 second to 15 seconds, with the average being around 3 seconds. The speech is recorded at 48 kHz. In training dataset, there are 10 different types of noise \cite{noise} added to the clean speech with 4 signal-to-noise (SNR) values: 15dB, 10dB, 5dB and 0dB. Thus the training dataset has 40 noisy conditions in total. In testing dataset, there are 5 types of noise which are added to the speech with 4 SNR values being 17.5dB, 12.5dB, 7.5dB and 2.5dB. The 28 speakers \cite{speakers} in training dataset are different from the 2 speakers \cite{speakers} in testing dataset, and all of them are native English speakers.
\vspace{-1.5mm}
\subsubsection{Metrics}
\vspace{-1mm}
We use five objective metrics to evaluate and compare the quality of the enhanced speech by the proposed coarse-to-fine optimization. \textbf{SSNR}, with the range from $-\infty$ to $\infty$, computes the segmental SNR in $dB$. \textbf{CSIG} and \textbf{CBAK} \cite{measure} respectively predict Mean Opinion Score (MOS) of the signal distortion attending to the speech signal alone and the background intrusiveness attending to the background noise alone. \textbf{COVL} \cite{measure} computes the MOS of the overall signal quality. \textbf{CSIG}, \textbf{CBAK} and \textbf{COVL} are all measured from 1 to 5. \textbf{PESQ} \cite{measure2}, standing for perceptual evaluation of the speech quality, is measured from $-0.5$ to $4.5$.
For all these metrics, the higher the measure is, the better quality the enhanced speech will have. As is known, there is no single objective measure that correlates perfect with subjective evaluations for different speech distortions\cite{loizou}. Therefore, we need to take all the above metrics into account when evaluating the speech quality (\textbf{SSNR} is included as widely used, though its correlation with overall speech quality is low\cite{measure}).
\begin{figure}[t]
  \centering
  \includegraphics[width=0.6\linewidth]{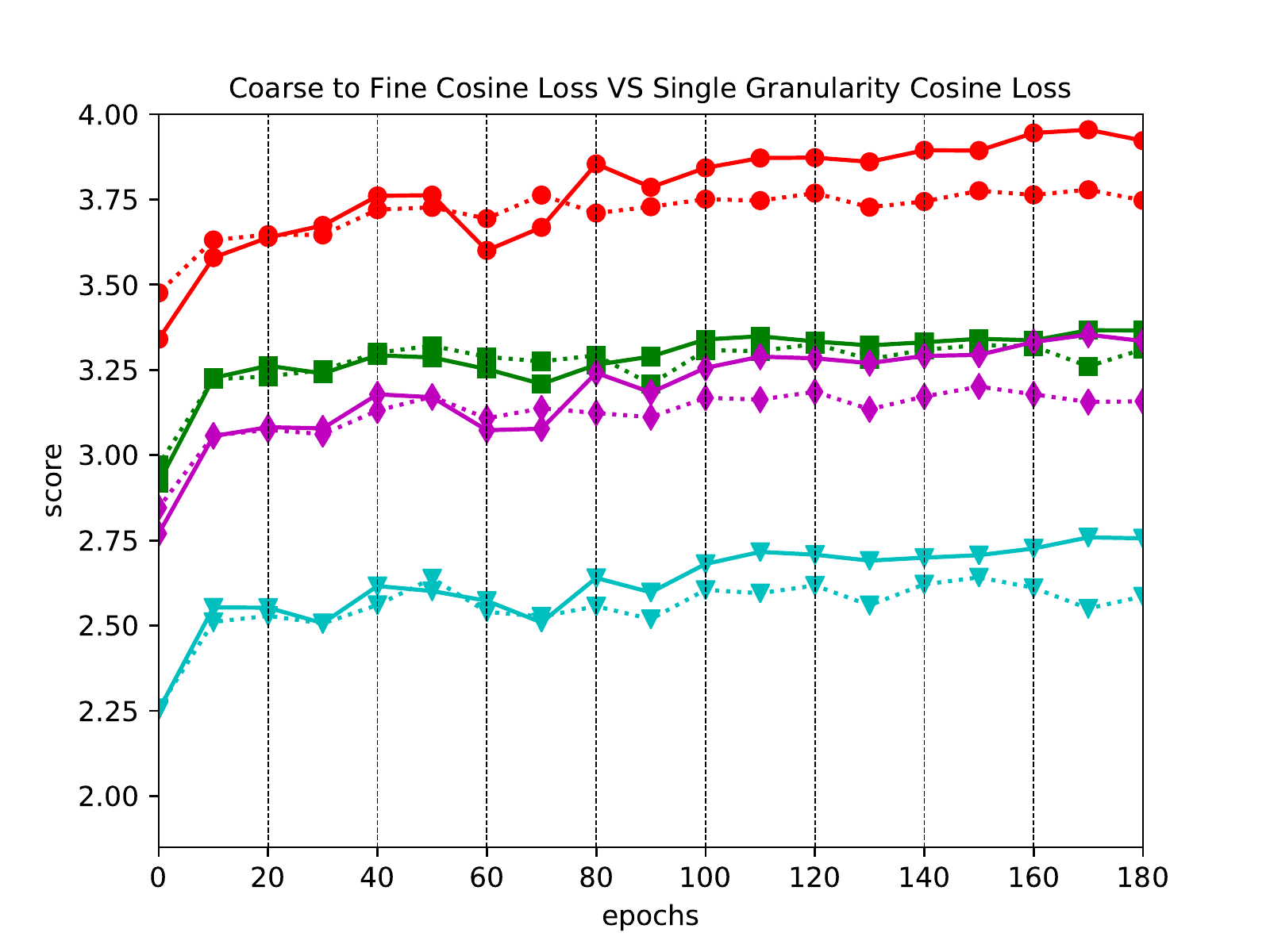}
  \vspace{-2mm}
  \caption{The coarse-to-fine optimization is applied to state-of-the-art discriminative model \cite{phase} and evaluated on test dataset. The solid line is the result from coarse-to-fine optimization and the dotted line is the result from optimizing the single-granularity cosine similarity loss. {\bf red:} CSIG, {\bf green:} CBAK, {\bf magenta:} COVL, {\bf cyan:} PESQ. \vspace{-6mm}}
  \label{fig:disc_coarse2fine}
\end{figure}
\subsection{Discriminative Model}
In this experiment, we compare our coarse-to-fine optimization with the state-of-the-art discriminative model \cite{phase} using single granularity optimization. 
Without loss of generality\footnote{The author of \cite{phase} confirmed there is technical error in the accuracy reported in the paper. The actual best accuracy from their method is similar to previous state-of-the-art methods (e.g. \cite{deepfeat}).}, we choose the network called cRM$\mathbb{R}$n in \cite{phase} to be the network in our discriminative model, as shown in Fig.~\ref{fig:network_arch_gnet}. The model is trained for 180 epochs with batch size 96 using Adam \cite{adam} optimizer. The initial learning rate is set to be 0.0004 and is multiplied by 0.5 at epoch 40, 80 and 120. The weight decay is 0.0005. We down-sample the input audio from 48 kHz to 16 kHz. And during training, similar to \cite{segan}, we divided the original audio into overlapped slices with the stride $2^{13}$, each of which has $2^{14}$ samples (approximately 1 second). During testing, as in \cite{segan}, we divide the test utterance into non-overlapped slices and concatenate the results as the final enhanced speech for the whole duration.

In the training, we compute the cosine similarity loss for both signal and background noise as well \cite{phase, wavenetSE} so that Eq.~\ref{eqncos} will be sensitive to the scale change of the signal. The granularity on which we compute our loss starts from the entire duration of the speech $2^{14}$ and decreases by 2 times every 20 epochs. The finest granularity is $2^6$. The vertical black dash line in Fig.~\ref{fig:disc_coarse2fine} indicates the moment we decrease the granularity by 2x. From the result shown in Fig.~\ref{fig:disc_coarse2fine}, we can see our coarse-to-fine optimization steadily outperforms the single-granularity optimization.
\begin{figure}[t]
  \centering
  \includegraphics[width=0.6\linewidth]{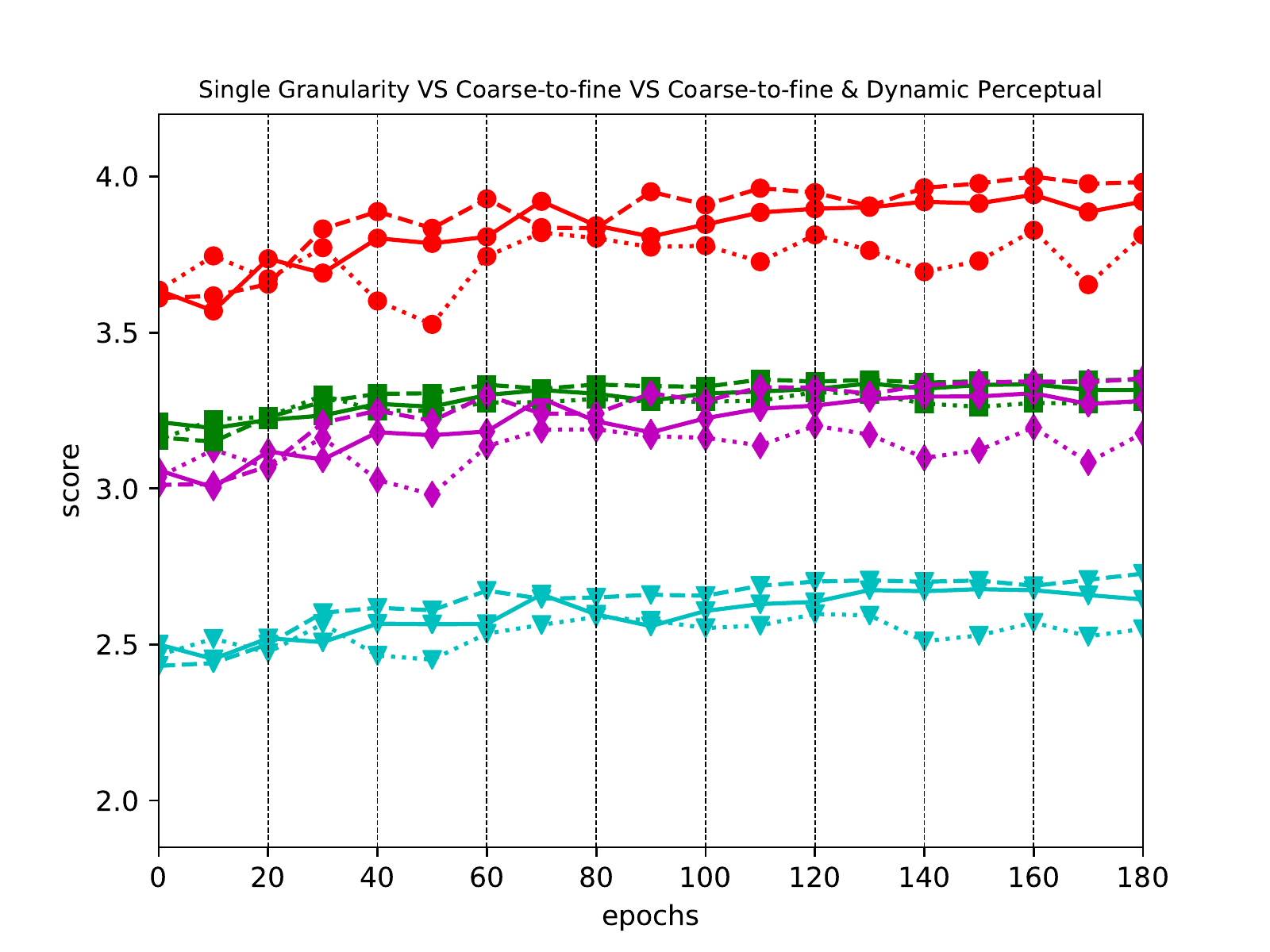}
  \vspace{-2mm}
  \caption{The comparison among \textbf{G}({\bf dotted line}), \textbf{G+M}({\bf solid line}) and \textbf{G+M+P}({\bf dashed line}) on test dataset. The color of lines represents the same metric as Fig.~\ref{fig:disc_coarse2fine}. \vspace{-6mm}}
  \label{fig:gan_comp}
\end{figure}
\subsection{Generative Model (GAN)}
In the GAN experiment, the batch size is reduced to 64 due to the memory limit. And the number of epochs for training is increased to 360 so that the effective number of epochs for the generator training is 180 as the discriminator and generator are trained alternatively one after the other \cite{segan}. The learning rate for the discriminator is constantly 0.0002. All other setups are the same as the discriminative model training. In the training of generator, the scalar coefficient for the regularization term is 40 while the coefficient for the discriminator adversarial loss is always 1. If the perceptual loss is considered, the balancing coefficient is 100. They are chosen in this way so that the scale of all the terms is almost the same. The regularization term for the generator is cosine similarity loss instead of $L1$ as widely used in other GAN methods\cite{segan, cmask}. We add a Gaussian noise with mean 0.0 and variance 0.01 between the encoder and the decoder of the generator.

As shown in Fig.~\ref{fig:gan_comp}, the plain GAN(\textbf{G}) means that the adversarial loss is the $L2$ norm as used in \cite{segan} and the regularization is cosine similarity loss. In coarse-to-fine optimization(\textbf{G+M}), we optimize the regularization in a coarse-to-fine way as used in the discriminative model. Furthermore, we apply the coarse-to-fine strategy to the adversarial loss term, i.e. dynamic perceptual loss(\textbf{G+M+P}). We start with the original $L2$ loss computed in the last layer of the discriminator as \textbf{G+M} and every 80 epochs, we compute the $L1$ loss on deep features of different resolution between the real and the fake utterance. The particular layers where the deep features are extracted are layer 9, 7, 5, 3. Fig.~\ref{fig:gan_comp} clearly shows \textbf{G+M+P} outperforms \textbf{G+M} which is better than \textbf{G}. 
In Tab.~\ref{tab:comp}, we compare our results with other popular speech enhancement algorithms. The accuracies of our models are always computed from the last iteration of the training without picking the best in histories. Tab.~\ref{tab:comp} further shows the effectiveness of our proposed coarse-to-fine optimization and the dynamic perceptual loss. The bold numbers highlight the best accuracy in either discriminative models or generative models, but not both. As shown in Tab.~\ref{tab:comp}, our discriminative model (\textbf{D+M}) and generative model (\textbf{G+M+D}) both outperform corresponding state-of-the-art methods in most metrics. We also notice that the single granularity in the discriminative model outperforms coarse-to-fine in SSNR. This can be explained by the consistent goal of measuring the difference over the entire speech between SSNR and single granularity loss, while not correlating well with overall speech quality compared with other metrics\cite{measure}.

%% file: conclusion.tex
\section{Conclusion and Discussion}
In this paper we proposed the coarse-to-fine strategy in optimizing the cosine similarity loss for both discriminative and generative models. Inspired by the coarse-to-fine idea, we further proposed the dynamic perceptual loss as the adversarial loss term in the generator training of GAN. Our experiments show the effectiveness of our proposed methods. In the future, we will look into the question of the better model choice for the task of speech enhancement: discriminative OR generative? And how to take both of their advantages into one model is still open. Dynamic perceptual loss might provide a direction, and it would also be very interesting to see whether it can be generalized to other applications besides speech enhancement.

%% file: main.bbl
\begin{thebibliography}{10}
\providecommand{\url}[1]{#1}
\csname url@samestyle\endcsname
\providecommand{\newblock}{\relax}
\providecommand{\bibinfo}[2]{#2}
\providecommand{\BIBentrySTDinterwordspacing}{\spaceskip=0pt\relax}
\providecommand{\BIBentryALTinterwordstretchfactor}{4}
\providecommand{\BIBentryALTinterwordspacing}{\spaceskip=\fontdimen2\font plus
\BIBentryALTinterwordstretchfactor\fontdimen3\font minus
  \fontdimen4\font\relax}
\providecommand{\BIBforeignlanguage}[2]{{%
\expandafter\ifx\csname l@#1\endcsname\relax
\typeout{** WARNING: IEEEtran.bst: No hyphenation pattern has been}%
\typeout{** loaded for the language `#1'. Using the pattern for}%
\typeout{** the default language instead.}%
\else
\language=\csname l@#1\endcsname
\fi
#2}}
\providecommand{\BIBdecl}{\relax}
\BIBdecl

\bibitem{cosineSim}
S.~Novoselov, V.~Shchemelinin, A.~Shulipa, A.~Kozlov, and I.~Kremnev, ``Triplet
  loss based cosine similarity metric learning for text-independent speaker
  recognition,'' in \emph{International Speech Communication Association
  Conference (Interspeech)}, 2018.

\bibitem{specsub}
M.~Berouti, R.~Schwartz, and J.~Makhoul, ``Enhancement of speech corrupted by
  acoustic noise,'' \emph{Proc. of the Int. Conf. on Acoustics, Speech, and
  Signal Processing (ICASS)}, vol.~4, pp. 208--211, 1979.

\bibitem{wiener}
J.~Lim and A.~Oppenheim, ``All-pole modeling of degraded speech,'' \emph{IEEE
  Trans. on Acoustics, Speech, and Signal Processing}, vol.~26, no.~3, pp.
  197--210, 1978.

\bibitem{segan}
S.~Pascual, A.~Bonafonte, and J.~Serra, ``Segan: Speech enhancement generative
  adversarial network,'' in \emph{International Speech Communication
  Association Conference (Interspeech)}, 2017.

\bibitem{deepfeat}
F.~G. Germain, Q.~Chen, and V.~Koltun, ``Speech denoising with deep feature
  losses,'' in \emph{https://arxiv.org/abs/1806.10522}, 2018.

\bibitem{mgan}
M.~H. Soni, N.~Shah, and H.~A. Patil, ``Time-frequency masking-based speech
  enhancement using generative adversarial network,'' in \emph{International
  Conference on Acoustics, Speech, and Signal Processing (ICASSP)}, 2018.

\bibitem{rnnse}
K.~Tan and D.~Wang, ``A convolutional recurrent neural network for real-time
  speech enhancement,'' in \emph{International Speech Communication Association
  Conference (Interspeech)}, 2018.

\bibitem{phase}
H.-S. Choi, J.-H. Kim, J.~Huh, A.~Kim, J.-W. Ha, and K.~Lee, ``Phase-aware
  speech enhancement with deep complex u-net,'' in \emph{International
  Conference on Learning Representations(ICLR)}, 2019.

\bibitem{waveform_utter}
S.~Fu, T.~Wang, Y.~Tsao, X.~Lu, and H.~Kawai, ``End-to-end waveform utterance
  enhancement for direct evaluation metrics optimization by fully convolutional
  neural networks,'' \emph{IEEE/ACM Trans. on Audio, Speech, and Language
  Processing}, vol.~26, no.~9, pp. 1570--1584, 2018.

\bibitem{add1}
D.~Michelsanti and Z.-H. Tan, ``Conditional generative adversarial networks for
  speech enhancement and noise-robust speaker verification,'' in
  \emph{International Speech Communication Association Conference
  (Interspeech)}, 2017.

\bibitem{add2}
T.~Kaneko, S.~Takaki, H.~Kameoka, and J.~Yamagishi, ``Generative adversarial
  network based postfilter for stft spectrograms,'' in \emph{International
  Speech Communication Association Conference (Interspeech)}, 2017.

\bibitem{reg}
Y.~Xu, J.~Du, L.-R. Dai, and C.-H. Lee, ``A regression approach to speech
  enhancement based on deep neural networks,'' in \emph{IEEE/ACM Trans. on
  Audio, Speech and Language Processing}, 2015.

\bibitem{multinoise}
A.~Kuman and D.~Florencio, ``Speech enhancement in multiple noise conditions
  using deep neural networks,'' in \emph{Int. Speech Communication Association
  Conf. (Interspeech)}, 2016.

\bibitem{lstmEN}
F.~Weninger, H.~Erdogan, S.~Watanabe, E.~Vincent, J.~L. Roux, J.~R. Hershey,
  and B.~Schuller, ``Speech enhancement with lstm recurrent neural networks and
  its application to noise-robust asr,'' in \emph{Int. Conf. on Latent Variable
  Analysis and Signal Separation}, 2015.

\bibitem{autoencoder}
X.~Lu, Y.~Tsao, S.~Matsuda, and C.~Hori, ``Speech enhancement based on deep
  denoising autoencoder,'' in \emph{International Speech Communication
  Association Conference (Interspeech)}, 2013.

\bibitem{gan}
I.~J. Goodfellow, J.~Pouget-Abadie, M.~Mirza, B.~Xu, D.~Warde-Farley, S.~Ozair,
  A.~Courville, and Y.~Bengio, ``Generative adversarial nets,'' in \emph{Nerual
  Information Processing (NIPS)}, 2014.

\bibitem{cgan}
M.~Mirza and S.~Osindero, ``Conditional generative adversarial nets,'' in
  \emph{https://arxiv.org/pdf/1411.1784.pdf}, 2014.

\bibitem{pixel2pixel}
P.~Isola, J.~Zhu, T.~Zhou, and A.~A. Efros, ``Image-to-image translation with
  conditional adversarial networks,'' in \emph{Conference of Computer Vision
  and Pattern Recognition(CVPR)}, 2017.

\bibitem{z1}
M.~Mathieu, C.~Couprie, and Y.~LeCun, ``Deep multi-scale video prediction
  beyond mean square error,'' in \emph{International Conference on Learning
  Representations(ICLR)}, 2016.

\bibitem{z2}
I.~J. Goodfellow, J.~Pouget-Abadie, M.~Mirza, B.~Xu, D.~Warde-Farley, S.~Ozair,
  A.~Courville, and Y.~Bengio, ``Generative image modeling using style and
  structure adversarial networks,'' in \emph{European Conference on Computer
  Vision(ECCV)}, 2016.

\bibitem{cnnse}
S.~R. Park and J.~W. Lee, ``A fully convolutional neural network for speech
  enhancement,'' in \emph{International Speech Communication Association
  Conference (Interspeech)}, 2017.

\bibitem{c2f1}
J.~Yao, M.~Boben, S.~Fidler, and R.~Urtasun, ``Real-time coarse-to-fine
  topologically preserving segmentation,'' in \emph{Conference of Computer
  Vision and Pattern Recognition (CVPR)}, 2015.

\bibitem{c2f2}
D.~J. Fleet and Y.~Weiss, ``Optical flow estimation,'' in
  \emph{http://www.cs.toronto.edu/~fleet/research/Papers/flowChapter05.pdf},
  2005.

\bibitem{stft}
J.~B. Allen, ``Short term spectral analysis, synthesis, and modification by
  discrete fourier transform,'' \emph{IEEE Transactions on Acoustics, Speech,
  Signal Processing}, vol. ASSP-25, pp. 235--238, 1977.

\bibitem{cmask}
D.~S. Williamson and D.~Wang, ``Time-frequency masking in the complex domain
  for speech dereverberation and denoising,'' \emph{IEEE/ACM TRANSACTIONS ON
  AUDIO, SPEECH, AND LANGUAGE PROCESSING}, vol.~25, no.~7, 2017.

\bibitem{deconv}
V.~Dumoulin and F.~Visin, ``A guide to convolution arithmetic for deep
  learning,'' in \emph{https://arxiv.org/pdf/1603.07285.pdf}, 2015.

\bibitem{bn}
S.~Ioffe and C.~Szegedy, ``Batch normalization: Accelerating deep network
  training by reducing internal covariate shift,'' in \emph{International
  Conference on Machine Learning (ICML)}, 2015, pp. 448--456.

\bibitem{lrelu}
A.~L. Maas, A.~Y. Hannun, and A.~Y. Ng, ``Rectifier nonlinearities improve
  neural network acoustic models,'' in \emph{International Conference on
  Machine Learning (ICML)}, 2013, p.~3.

\bibitem{deepface}
Y.~Taigman, M.~Yang, M.~A. Ranzato, and L.~Wolf, ``Deepface: Closing the gap to
  human-level performance in face verification,'' in \emph{Conference of
  Computer Vision and Pattern Recognition(CVPR)}, 2014.

\bibitem{wavenetSE}
D.~Rethage, J.~Pons, and X.~Serra, ``A wavenet for speech denoising,'' in
  \emph{International Conference on Acoustics, Speech, and Signal Processing
  (ICASSP)}, 2018.

\bibitem{dataset}
C.~Valentini-Botinhao, ``Noisy speech database for training speech enhancement
  algorithms and tts models,'' in \emph{University of Edinburgh. School of
  Informatics. Centre fro Speech Technology Research (CSTR)
  https://doi.org/10.7488/ds/1356}, 2016.

\bibitem{noise}
J.~Thiemann, N.~Ito, and E.~Vincent, ``The diverse environments multi-channel
  acoustic noise database: A database of multichannel environmental noise
  recordings,'' \emph{The Journal of the Acoustical Society of America}, vol.
  133, no.~5, pp. 3591--3591, 2013.

\bibitem{speakers}
C.~Veaux, J.~Yamagishi, and S.~King, ``The voice bank corpus: Design,
  collection and data analysis of a large regional accent speech database,'' in
  \emph{Int. Conf. Oriental COCOSDA, held jointly with 2013 Conference on Asian
  Spoken Language Research and Evaluation (O-COCOSDA/CASLRE)}, 2013, pp. 1--4.

\bibitem{measure}
Y.~Hu and P.~C. Loizou, ``Evaluation of objective quality measures for speech
  enhancement,'' \emph{IEEE Trans. on Audio, Speech and Language Processing},
  vol.~16, no.~1, pp. 229--238, 2008.

\bibitem{measure2}
``P.862.2: Wideband extension to recommendation p. 862 for the assessment of
  wideband telephone networks and speech codecs,'' in \emph{Geneva:
  International Telecommunication Union}, 2007.

\bibitem{loizou}
P.~C. Loizou, \emph{Speech Enhancement: Theory and Practice}, 2nd~ed.\hskip 1em
  plus 0.5em minus 0.4em\relax Boca Raton, FL, USA: CRC Press, 2013.

\bibitem{adam}
D.~P. Kingma and J.~Ba, ``Adam: A method for stochastic optimization,'' in
  \emph{International Conference for Learning Representation (ICLR)}, 2015.

\end{thebibliography}
